\documentclass[10pt,english]{article}
\usepackage{mathptmx}
\usepackage[T1]{fontenc}
\usepackage[latin9]{luainputenc}
\usepackage{xcolor}
\usepackage{babel}
\usepackage{amsmath}
\usepackage{amsthm}
\usepackage{amssymb}
\usepackage{setspace}
\PassOptionsToPackage{normalem}{ulem}
\usepackage{ulem}
\usepackage[pdfusetitle,
 bookmarks=true,bookmarksnumbered=false,bookmarksopen=false,
 breaklinks=false,pdfborder={0 0 1},backref=false,colorlinks=false]
 {hyperref}

\makeatletter

\providecolor{lyxadded}{rgb}{0,0,1}
\providecolor{lyxdeleted}{rgb}{1,0,0}
\DeclareRobustCommand{\mklyxadded}[1]{\bgroup\color{lyxadded}{}#1\egroup}
\DeclareRobustCommand{\mklyxdeleted}[1]{\bgroup\color{lyxdeleted}\mklyxsout{#1}\egroup}
\DeclareRobustCommand{\mklyxsout}[1]{\ifx\\#1\else\sout{#1}\fi}

\numberwithin{figure}{section}
\newcommand{\lyxaddress}[1]{
	\par {\raggedright #1
	\vspace{1.4em}
	\noindent\par}
}

\makeatother

\begin{document}
\title{\noindent On the Pure States of the Replica Symmetry Breaking ansatz}
\author{\noindent Simone Franchini}
\date{~}
\maketitle

\lyxaddress{\begin{center}
\textit{Goethe University Mathematics Institute, }\\
\textit{10 Robert Mayer }\textit{Str}\textit{, 60325 Frankfurt, Germany}
\par\end{center}}
\begin{abstract}
\noindent We discuss the concept of Pure State of the Replica Symmetry
Breaking ansatz in finite and infinite spin systems without averaging
on the disorder, nor using replicas. Consider a system of $n$ spins
$\sigma\in\Omega^{n}$ with the usual set $\Omega=\left\{ -1,1\right\} $
of inner states and let $G:\,\Omega^{n}\rightarrow\left[0,1\right]$
a Gibbs measure on it of Hamiltonian $\mathcal{H}$ (also non random).
We interpret the pure states of a model $\left(\Omega^{n},\mu\right)$
as disjoint subsets $\Omega^{n}$ such that the conditional measures
behaves like product measures as in usual mean field approximations.
Starting from such definition we try to reinterpret the RSB scheme
and define an approximated probability measure. We then apply our
results to the Sherrington-Kirkpatrick model to obtain the Parisi
formula.

\noindent ~

\noindent ~
\end{abstract}

\section{Introduction}

Originally introduced by Parisi in order to interpret its solution
to the Sherrington-Kirkpatrick model (SK) \cite{Parisi-1,Parisi-2},
the Replica Symmetry Breaking (RSB) ansatz proved to be an extremely
valuable tool in explaining properties of disordered systems. Despite
many technical advances, worth to cite the proof of the free energy
functional by Guerra and Talagrand \cite{Guerra,Talagarand}, some
of its key physical features remain quite mysterious after more than
thirty years. 

A central role is played by the elusive concept of \textquotedblleft pure
state\textquotedblright . Despite a precise definition is still lacking,
it is widely acknowledged that they must satisfy some properties.
As example, it is expected that the connected correlation functions
conditioned to these subsets vanishes in the thermodynamic limit \cite{Parisi-2}.
This imply that in some sense the measure conditioned to those states
can be described by a mean field model.

Perhaps, the most striking and unconventional property is that the
pure states are predicted to have a hierarchical structure such that
the support of the overlap is ultra- metric \cite{Parisi-2}. A considerable
amount of work has been produced on this subject, culminating in a
proof of ultrametricity for the SK model by Panchenko \cite{Panchenko}.
Anyway, if ultrametricity and other properties of the pure states
hold in some general framework, including their very existence as
well defined mathematical objects, proved to be an extremely hard
task and remains an open question.

A common assumption in almost all the above literature is that pure
states are expected to represents thermodynamic phases, thus being
collections of a thermodynamically relevant number of samples. Inspired
by a recent work which connects graph theory and Belief Propagation
\cite{ACO-1}, we propose that the set of pure states of the RSB ansatz
can be represented as a partition of $\Omega_{n}$ into subsets $S=\left\{ S_{i}\right\} $
such that the probability measure conditioned on each Si can be described
by a system of non interacting spins coupled to a non homogeneous
external field. As we shall see, this introduces critical simplifications
in reproducing the results of the RSB scheme, which we interpret as
a technique to approximate random Gibbs measures trough a weighted
\textquotedblleft mixtures\textquotedblright{} of mean field models.

The paper is organized as follows. In the first section we will discuss
the concept of Regularity Partitions for separately exchangeable arrays.
Then we will describe how to construct a probability measure which
we expect to be equivalent to the Replica Symmetric approximation
and then generalize this argument to obtain any finite RSB measure.
Finally, we apply this measure to the Aizenmann-Simm-Starr representation
for the free energy of the SK model and derive the Parisi functional.

\section{Regularity partitions.}

Before entering in the core of the discussion, a little mathematical
digression is mandatory in order to justify our later arguments.

Let $\Omega^{n}$ the product space of $n$ spins with finite number
of inner states, let $\mathcal{P}\left(\Omega^{n}\right)$ the ensemble
of all probability measures on $\Omega^{n}$ and let $\mathcal{\mu\in P}\left(\Omega^{n}\right)$
some probability measure. We denote by $\left(\Omega^{n},\mu\right)$
our model and by $\mu_{K}$ the marginal distribution of $\mu$ over
a subset $K\subset\left\{ 1\,...\,n\right\} $ of coordinates. If
$S=\left\{ S_{i}\right\} $, $i\in\{1\,...\,\left|S\right|\}$ is
a partition of $\Omega^{n}$ into $\left|S\right|$ disjoint subsets
we call $\mu^{i}$ the measure conditioned to $S_{i}$ and by $\mu_{K}^{i}$
the marginal distribution of $\mu^{i}$ over $K$. The connection
between graph theory and the RSB scheme has been first observed by
Coja-Oghlan et al. in \cite{ACO-1}, where it is shown that for any
measure $\mu$ on $\Omega^{n}$ it is possible take some arbitrarily
small $\epsilon>0$ and a partition $\Omega^{n}$ into a finite number
(not depending on $n$) of disjoint $\left\{ S_{i}\right\} ,$ $0\leq i\leq\left|S\right|$
such that $\mu\left(S_{0}\right)\leq\epsilon$ and the marginals factorize
\begin{equation}
\sum_{K\in\left\{ 1,\,...\,n\right\} ^{\left|K\right|}}\left\Vert {\textstyle \mu_{K}^{i}-\bigotimes_{\alpha\in K}\mu_{\alpha}^{i}}\right\Vert _{TV}\leq\epsilon\,n^{\,\left|K\right|}\label{eq:DECOMP1}
\end{equation}
if $n$ is chosen large enough (we denoted by $\left\Vert \,\cdot\,\right\Vert _{TV}$
the total variation distance). The above result tell us that for any
measure on a system of variables with finite number of inner states
(here we assume $\Omega=\left\{ -1,1\right\} $) we can decompose
our sample space into a finite number of ``regular'' disjoint subsets
$\left\{ S_{i}\right\} $, $i\geq1$ plus one ``irregular'' $S_{0}$
with $\mu\left(S_{0}\right)\leq\epsilon$ such that for any regular
subset $S_{i}$ the marginals of $\mu^{i}$ over a randomly chosen
set $K$ can be approximated by a product measure in the sense of
Eq. (\ref{eq:DECOMP1}). The number of such regular subsets only depends
on $\left|K\right|$, $\left|\Omega\right|$ and the level of precision
$\epsilon$ we want to achieve for our approximation, and it does
not depend on the size $n$ of the system. This implies that at least
for some $S_{i}$ we have $\mu\left(S_{i}\right)>0$ and we are allowed
interpret them as thermodynamic phases when the $n\rightarrow\infty$
limit is taken.

The logic behind the above result lies on a graph theoretic argument
known as Szemerédi Regularity lemma, which in one its versions says
that any kernel (ie, a bounded measurable function $\left[0,1\right)\times\left[0,1\right)\rightarrow\left(-\infty,\infty\right)$)
can be approximated almost everywhere in $L_{1}$ norm by a step function
with finite number of steps. 

From an intuitive point of view this lemma provides an extension to
distributions (ie, Lebesgue integrable functions) of the fact that
any continuous function can be arbitrarily approximated by a step
function in $L_{1}$ norm. Indeed, it can be seen as a Riemann integrable
approximation of a Lebesgue integrable function. More formally, let
us label the spin vectors of our sample space $\Omega^{n}$ as follows
\begin{equation}
\Omega^{n}=\left\{ \sigma^{k}\right\} _{1\leq k\leq2^{n}},\,\sigma^{k}=\left\{ \sigma_{a}^{k}\right\} _{1\leq a\leq n}
\end{equation}
and define the exchangeable array 
\begin{equation}
\mathcal{M}=\left\{ M_{a}^{k}\right\} _{1\leq a\leq n,\,1\leq k\leq2^{n}}:\,M_{a}^{k}=\mu\left(\sigma^{k}\right)\sigma_{a}^{k}.\label{MagnetizationKer}
\end{equation}
For some interval $A\times B\subseteq\left\{ 1,\,...\,n\right\} \times\Omega^{n}$
denote the mean value of $M$ by 
\begin{equation}
\overline{\mathcal{M}}\left(A,B\right)=\frac{1}{\left|A\right|\left|B\right|}\sum_{a\in A,\,k\in B}M_{a}^{k}=\sum_{a\in A}\langle\,\sigma_{a}^{k}\,\mathbb{I}_{\left\{ \sigma^{k}\in B\right\} }\rangle_{\mu}.
\end{equation}
In \cite{ACO-1} it is shown that for any $\epsilon>0$ exists a pair
of irregular intervals $\left(V_{0},S_{0}\right)\subset(\left\{ 1,\,...\,n\right\} ,\Omega^{n})$
with $\left|V_{0}\right|<\epsilon n$, $\left|S_{0}\right|<\epsilon\left|\Omega^{n}\right|$
and a pair of regular partitions
\begin{equation}
\left(V,S\right)=(\,\left\{ V_{\alpha}\right\} _{1\leq\alpha\leq\left|V\right|},\,\left\{ S_{i}\right\} _{1\leq i\leq\left|S\right|}\,)
\end{equation}
of $\left\{ 1,\,...\,n\right\} \setminus V_{0}\times\left[0,1\right)\setminus S_{0}$
into a finite number of sub intervals such that for any $A\times B\subseteq V_{\alpha}\times S_{i}$
with $\left|A\right|\geq\epsilon\left|V_{\alpha}\right|$, $\left|B\right|\geq\epsilon\left|S_{i}\right|$
holds that $\left|\overline{\mathcal{M}}\left(A,B\right)-\overline{\mathcal{M}}\left(V_{\alpha},S{}_{i}\right)\right|\leq\epsilon$
if $n$ is taken large enough, ie if $n\geq n^{*}\left(\epsilon,\Omega\right)$
where $n^{*}\left(\epsilon,\Omega\right)<\infty$ does not depend
on $n$. Eq. (\ref{eq:DECOMP1}) almost immediately follows from noticing
that if we take $\left|K\right|$ points randomly in $\left\{ 1,\,...\,,n\right\} $
they will be contained in the regular intervals $V$ with probability
$\bar{\epsilon}^{\,\left|K\right|}$, $\bar{\epsilon}=1-\epsilon$
(hereafter for any number $c\in\left[0,1\right]$ we denote its complement
$\bar{c}=1-c$ with an overbar). A formal proof of Eq. (\ref{eq:DECOMP1})
can be found in \cite{ACO-1}, but we stress that it will become intuitively
evident in the next paragraph, when we introduce our approximated
model.

Before going ahead we state the above result in a continuous form,
ie in a kernel form \cite{Lovasz}, so that we can use a unified notation
both for $n<\infty$ and $n\rightarrow\infty$. We define a ``magnetization
kernel'' $W:\left[0,1\right)\times\left[0,1\right)\rightarrow\left[-1,1\right]$
associated to $\left(\Omega^{n},\mu\right)$ as 
\begin{equation}
W\left(x,y\right)=\sum_{\alpha=1}^{n}\,\sum_{k=1}^{2^{n}}\sigma_{a}^{k}\,\mathbb{I}_{\left\{ \left(x,y\right)\in\left[x_{a-1},x_{a}\right)\times\left[y_{k-1},y_{k}\right)\right\} }
\end{equation}
where $x_{a}=a/n$ and $y_{k}=\sum_{j=1}^{k}\mu(\sigma^{j})$. Then,
for some interval $A\times B\subseteq\left[0,1\right)\times\left[0,1\right)$
let denote the mean value of $W$ by 
\begin{equation}
\overline{W}\left(A,B\right)=\frac{1}{\left|A\right|\left|B\right|}\int_{A\times B}W\left(x,y\right)dx\,dy,
\end{equation}
Since $W$ is a kernel, by Szemerédi Regularity Lemma \cite{Lovasz}
for any $\epsilon>0$ there exist a pair of irregular intervals $\left(V_{0},\Sigma_{0}\right)\subset(\left[0,1\right),\left[0,1\right))$
with $\left|V_{0}\right|<\epsilon$, $\left|\Sigma_{0}\right|<\epsilon$
and a pair of $\epsilon-$regular partitions
\begin{equation}
\left(V,\Sigma\right)=(\,\left\{ V_{\alpha}\right\} _{1\leq\alpha\leq\left|V\right|},\,\left\{ \Sigma_{i}\right\} _{1\leq i\leq\left|\Sigma\right|}\,)
\end{equation}
of $\left[0,1\right)\setminus V_{0}\times\left[0,1\right)\setminus S_{0}$
into a finite number of subintervals such that for any $A\times B\subseteq V_{\alpha}\times\Sigma_{i}$
with $\left|A\right|\geq\epsilon\left|V_{\alpha}\right|$, $\left|B\right|\geq\epsilon\left|\Sigma_{i}\right|$
holds that $\left|\overline{W}\left(A,B\right)-\overline{W}\left(V_{\alpha},\Sigma{}_{i}\right)\right|\leq\epsilon$.
A proof can be found in Chapter 9 of \cite{Lovasz}. We will call
by $\varphi$ the map that keep track of the correspondence $\varphi\Sigma_{i}=S_{i}$,
$\Sigma_{i}=\varphi^{-1}S_{i}$ and $\left|\Sigma_{i}\right|=\left|\varphi^{-1}S_{i}\right|=\mu\left(S_{i}\right)$. 

Clearly the above statement is nontrivial only if $\overline{W}\left(V_{\alpha},\Sigma{}_{i}\right)=O\left(1\right)$,
which is true only if at east some spin configurations carry a finite
fraction of the probability mass as $n\rightarrow\infty$, ie the
number of pure state is numerable or countably infinite. Since we
are here interested in models that exhibit the fullRSB picture, we
will mainly work under the assumption that $\mu\left(\sigma^{k}\right)\leq c_{n}$,
$c_{n}\rightarrow\infty$ in $n$ such that no single configuration
carries a finite probability mass in the thermodynamic limit. By assuming
$\mu\left(\sigma^{k}\right)=o\left(1\right)$ and $\left|V_{\alpha}\right|=O\left(1\right)$
we need to take $|\Sigma|$ to be increasing in $n$ in order to ensure
that $\overline{W}\left(V_{\alpha},\Sigma{}_{i}\right)=O\left(1\right)$. 

It is also crucial to notice that any refinement of an $\epsilon-$regular
partition $\left(V,S\right)$, ie a partition generated by splitting
each $(V_{\alpha},\Sigma_{i})$, produces again an $\epsilon-$regular
partition, in the sense that if $\left(V',\Sigma'\right)$ is a refinement
of $\left(V,\Sigma\right)$ then the error $\epsilon'$ associated
to the refinement is $\epsilon'<\epsilon$. Then, any refinement of
the set of pure states is again a set of pure states, and given a
partition $\left(V,S\right)$ we can produce an equitable version
$(V^{'},\Sigma^{'})$ (ie, where $|V_{\alpha}^{'}|$ and $|\Sigma_{i}^{'}|$
are of equal sizes) which has at most the same $\epsilon$. 

Combined to the fact that we can chose $\epsilon$ arbitrary small
if $n$ is arbitrarily large this provides a compactness argument
for the magnetization kernel space \cite{Lovasz} and we can equivalently
consider $\left(V,\Sigma\right)$ to be an equitable partition of
the kernel $W$ with 
\begin{equation}
\left|V_{\alpha}\right|=1/N_{V}>0,\ \left|\Sigma_{i}\right|=1/N_{\Sigma}=|\Omega|^{(1/N_{V}-1)n}
\end{equation}
for all $\alpha$ and $i$. The $n-$dependent scaling between $\left|\Sigma_{i}\right|$
and $\left|V_{\alpha}\right|$ has been taken in order to allow both
$\overline{W}\left(V_{\alpha},\Sigma{}_{i}\right)=O\left(1\right)$
in the $n\rightarrow\infty$ limit while keeping $\left|V_{\alpha}\right|=O\left(1\right)$.
The choice of the ratio constant $\log\left|\Omega\right|$ is uninfluent.

\section{A mixture of mean-field approximations }

As we shall see, these regularity partitions can be tributed of a
physical interpretation in therms of a collection of subsystems which
behaves approximately as non interacting spins coupled to spatially
non homogeneous external fields. The general structure is described
by the array 
\begin{equation}
T=\left\{ \tau^{i}\right\} _{0\leq i\leq N_{\Sigma}},\ \tau^{i}=\left\{ \tau_{a}^{i}\right\} _{0\leq a\leq n},\ \tau_{a}^{i}\in\left[-1,1\right].\label{eq:TAU}
\end{equation}
from which our approximating models are constructed by taking
\begin{equation}
\eta\left(\sigma\right)=N_{\Sigma}^{-1}\sum_{i=1}^{N_{\Sigma}}\eta^{i}\left(\sigma\right),\ \eta^{i}\left(\sigma\right)=\prod_{a=0}^{n}\left(\frac{1+\tau_{a}^{i}\sigma_{a}}{2}\right).\label{eq:TAUMODEL}
\end{equation}
Then, $\left(\Omega^{n},\eta\right)$ simply describes a uniformly
weighted mixtures of non interacting systems coupled to external fields
$\tau^{i}$ determined by the choice of the array $T$. As we shall
see, in the case of the SK model the $h^{i}$ will turn to be just
the cavity fields. Now, let $\left(V,\Sigma\right)$ an equitable
partition with $|V_{\alpha}|=1/N_{V}$, $|\Sigma_{i}|=1/N_{\Sigma}$
and $N_{\Sigma}=|\Omega|^{(1-1/N_{V})n}$, let $W$ the magnetization
kernel for the model $\left(\Omega,\mu\right)$ and chose $T$ such
that 
\begin{equation}
\tau_{a}^{i}=\sum_{\alpha=1}^{N_{V}}m_{\alpha}^{i}\mathbb{I}_{\left\{ a\in V_{\alpha}\right\} },\ m_{\alpha}^{i}=\overline{W}\left(V_{\alpha},\Sigma^{i}\right),\label{eq:TAUker}
\end{equation}
We can safely take the rows and columns of the irregular sets to be
$0$, ie $\tau_{0}^{i}=0$ for all $a\in V_{0}$ and $\tau_{a}^{0}=0$
for all $a\in\left\{ 1,\,...\,,n\right\} $. By definition, this model
approximate the original one in the sense that its magnetization kernel
match the regular part of original one, hence giving $\left\Vert \mu-\eta\right\Vert _{TV}\leq\epsilon$
a.s. for fixed $\epsilon$ and $n$ large enough. 

The intuitive sense of this approximation is that we stop keeping
track of fluctuations if these are of order $\epsilon$. If we refine
enough our kernel we can substitute $\mu_{V_{\alpha}}^{i}$by an effective
product measure of i.i.d. Bernoulli variables of parameter $m_{\alpha}^{i}$
that match the mean value only of the actual distribution, all by
paying a price $\epsilon$ which can be made arbitrarily small. To
be more explicit, let start from the exact formula
\begin{equation}
\mu\left(\sigma\right)=\sum_{\sigma'\in\Omega^{n}}\mu\left(\sigma'\right)\prod_{a=1}^{n}\left(\frac{1+\sigma'_{a}\sigma_{a}}{2}\right)
\end{equation}
and rewrite it as follows
\begin{equation}
\mu\left(\sigma\right)=\sum_{i=1}^{N_{\Sigma}}\mu\left(S^{i}\right)\sum_{\sigma'\in S^{i}}\frac{\mu\left(\sigma'\right)}{\mu\left(S^{i}\right)}\prod_{a=1}^{n}\left(\frac{1+\sigma'_{a}\sigma_{a}}{2}\right)=\sum_{i=1}^{N_{\Sigma}}\mu\left(S^{i}\right)\mu^{i}\left(\sigma\right),\label{eq:fundamental}
\end{equation}
where $\mu^{i}\left(\sigma\right)$ is the measure conditioned to
$S^{i}$ 
\begin{equation}
\mu^{i}\left(\sigma\right)=\sum_{\sigma'\in S^{i}}\frac{\mu\left(\sigma'\right)}{\mu\left(S^{i}\right)}\prod_{a=1}^{n}\left(\frac{1+\sigma'_{a}\sigma_{a}}{2}\right).
\end{equation}
The approximation consists in replacing $\mu^{i}\left(\sigma\right)$
with the product measure 
\begin{equation}
\mu^{i}\left(\sigma\right)\rightarrow\eta^{i}\left(\sigma\right)=\prod_{a=1}^{n}\left(\frac{1+\tau_{a}^{i}\sigma_{a}}{2}\right)
\end{equation}
and argue that $\mu\left(\sigma\right)=\eta\left(\sigma\right)+O\left(\epsilon\right)$
by means of the kernel argument presented in the first section. 
\begin{multline}
\eta\left(\sigma\right)=\sum_{i=1}^{N_{\Sigma}}\mu\left(S^{i}\right)\prod_{a=1}^{n}\left(\frac{1+\tau_{a}^{i}\sigma_{a}}{2}\right)=\\
=\sum_{i=1}^{N_{\Sigma}}\mu\left(S^{i}\right)\prod_{\alpha=1}^{N_{V}}\prod_{a\in V_{\alpha}1}\left(\frac{1+\tau_{a}^{i}\sigma_{a}}{2}\right)=\sum_{i=1}^{N_{\Sigma}}\mu\left(S^{i}\right)\prod_{\alpha=1}^{N_{V}}\eta_{\alpha}^{i}\left(\sigma\right),\label{eq:fundamental-1}
\end{multline}
If we chose to work with an equitable partition we can take $\mu\left(S^{i}\right)=1/N_{\Sigma}$
and 
\begin{equation}
\mu\left(\sigma\right)=N_{\Sigma}^{-1}\sum_{i=1}^{N_{\Sigma}}\prod_{\alpha=1}^{N_{V}}\eta_{\alpha}^{i}\left(\sigma\right)+O\left(\epsilon\right).
\end{equation}
Since the array $\{m_{\alpha}^{i}\}$ is separately exchangeable we
can perform a random mixing of the indexes $\alpha$, $i$ and think
$m_{\alpha}^{i}$ as $N_{\Sigma}$ random vectors extracted according
to some law $\zeta$. In the following of this paper we denote with
$\mathbf{m}=\{\mathbf{m}_{\alpha}\}$ the random vector representing
the magnetization kernel, and the average respect to $\mathbf{m}\sim\zeta$
as
\begin{equation}
N_{\Sigma}^{-1}\sum_{i=1}^{N_{\Sigma}}\left(\,\cdot\,\right)\rightarrow\mathbb{E}_{\,\mathbf{m}_{\alpha}}\left(\,\cdot\,\right),
\end{equation}
where the index $i$ is now dropped. Then, we will often express the
approximated measure $\eta\left(\sigma\right)$ using the notation
\begin{equation}
\eta\left(\sigma\right)=\mathbb{E}_{\,\mathbf{m}_{\alpha}}\prod_{\alpha=1}^{N_{V}}\boldsymbol{\eta}_{\alpha}\left(\sigma\right)\label{eq:aproximantmes}
\end{equation}
where $\boldsymbol{\eta}_{\alpha}\left(\sigma\right)$ is just $\eta_{\alpha}^{i}\left(\sigma\right)$
with $\mathbf{m}_{\alpha}$ on behalf of $m_{\alpha}^{i}$ in the
definition, and the dependence on the $\mathbf{m}_{\alpha}$ random
variables is kept implicit. We will also denote the averages with
respect to the measure $\mu$ as
\begin{equation}
\sum_{\sigma'\in\Omega^{n}}\mu\left(\sigma'\right)\left(\,\cdot\,\right)\rightarrow\langle\,\cdot\,\rangle_{\mu},
\end{equation}
Here comes the first interesting observation. Notice that $|V_{\alpha}|=O\left(n\right)$
and $\log|\Sigma_{i}|=O\left(n\right)$, and we may be tempted to
invoke the Doob's representation and the Central Limit Theorem to
approximate the law $\zeta$ with a normal distribution $N\left(m,\gamma\right)$,
ie approximate in law $\mathbf{m}$ with a Gaussian vector of i.i.d.
random variables 
\begin{equation}
\mathbf{m}_{\alpha}\overset{\mathcal{L}}{=}m+\mathbf{z}_{\alpha}\sqrt{\gamma},\ \mathbf{z}_{\alpha}\sim N\left(0,1\right).
\end{equation}
where $\overset{\mathcal{L}}{=}$ indicates equality in law (ie, in
distribution). If we do this we obtain an approximated probability
measure 
\begin{equation}
\eta_{RS}\left(\sigma\right)=\mathbb{E}_{\,\mathbf{z}_{\alpha}}\prod_{\alpha=1}^{N_{V}}\prod_{a\in V_{\alpha}}\left[\frac{1+\left(m+\mathbf{z}_{\alpha}\sqrt{\gamma}\right)\sigma_{a}}{2}\right]
\end{equation}
which only depend on the parameters $m=\overline{W}\left(V^{0},\Sigma^{0}\right)$
and $\gamma=\mathbb{E}_{\mathbf{m}_{\alpha}}|\mathbf{m}_{\alpha}-m|^{2}$
empirical estimator of the variance between the blocks. We expect
the above to be equivalent to the Replica Symmetric (RS) approximation,
we will further discuss this in the next section. 

Since we will concentrate our attention on Gibbs measures, before
going ahead it will be useful to restate Eq. (\ref{eq:TAUMODEL})
in better shape by introducing the ``cavity'' kernel
\begin{equation}
V=\left\{ \nu^{i}\right\} _{0\leq i\leq N_{\Sigma}},\ \nu^{i}=\left\{ \nu_{a}^{i}\right\} _{1\leq a\leq n}\label{eq:CAVITY}
\end{equation}
related to the magnetization kernel by the following definitions
\begin{equation}
\nu_{a}^{i}=\sum_{\alpha=1}^{N_{V}}h_{\alpha}^{i}\mathbb{I}_{\left\{ a\in V_{\alpha}\right\} },\ h_{\alpha}^{i}=\tanh^{-1}\left(m_{\alpha}^{i}\right).\label{eq:TAUker-1}
\end{equation}
Then, the $\eta_{\alpha}^{i}$ can be represented as 
\begin{equation}
\eta_{\alpha}^{i}\left(\sigma\right)=\prod_{a\in V_{\alpha}}\frac{1}{Z_{\alpha}^{i}}\,e^{-h_{\alpha}^{i}\sigma_{a}}
\end{equation}
where the normalization $Z_{\alpha}^{i}$ is given by
\begin{equation}
\log Z_{\alpha}^{i}=\log2+\log\cosh\left(h_{\alpha}^{i}\right).
\end{equation}
We can immediately recognize the partition of $\Omega^{n}$ as a partition
into subsets of configurations which linearize the Hamiltonian operator
$\mathcal{H}$, ie such that for any $\sigma\in S_{i}$ we approximate
$\mathcal{H}\left(\sigma\right)\rightarrow\nu^{i}\sigma$ as in the
usual mean field theories. Again, we can express the $\{h_{\alpha}^{i}\}$
as random vectors $\boldsymbol{h}$ extracted according to the law
$\xi\leftarrow\zeta$ and write for $\eta\left(\sigma\right)$ the
equivalent representation 
\begin{equation}
\eta\left(\sigma\right)=\mathbb{E}_{\,\mathbf{h_{\alpha}}}\prod_{\alpha=1}^{N_{V}}\boldsymbol{\eta}_{\alpha}\left(\sigma\right)
\end{equation}
Before turning on the implications of the the above in the interpretation
of the RSB scheme we recall that the free energy $F$ is defined as
$F=\log Z$ but we can give an alternative definition by the functional
\begin{equation}
\mathcal{F}_{\beta}\left[\mu,\mathcal{H}\right]=\langle\mathcal{H}\left(\sigma\right)-\frac{1}{\beta}\log\mu\left(\sigma\right)\rangle_{\mu}.
\end{equation}
Then, the free energy is given by 
\begin{equation}
F=\mathcal{F}_{\beta}\left[G,\mathcal{H}\right]=\inf_{\mu\in\mathcal{P}\left(\Omega^{n}\right)}\mathcal{F}_{\beta}\left[\mu,\mathcal{H}\right],
\end{equation}
where $\mathcal{P}\left(\Omega^{n}\right)$ the set of probability
measures on $\Omega^{n}$. Since any measure $\mathcal{\mu\in P}\left(\Omega^{n}\right)$
can be arbitrary approximated by the mixture of product measures above
at the cost of an error $\epsilon$ in total variation, we are interested
in searching our infimum over the parameter set instead of $\mathcal{P}\left(\Omega^{n}\right)$.
Hence we also define
\begin{equation}
F^{*}=\mathcal{F}_{\beta}\left[G^{*},\mathcal{H}\right]=\inf_{T}\mathcal{F}_{\beta}\left[\eta,\mathcal{H}\right],
\end{equation}
where we minimize over the kernel $T$ in Eq. (\ref{eq:CAVITY}).
It will be also useful to give an expression for the overlap distribution
of the approximated model. Let $\sigma,\sigma'\in\Omega^{n}$ and
denote the overlap distribution of the $\left(\Omega^{n},\eta\right)$
model as $P\left(q\right)=\eta\otimes\eta\left(\sigma\sigma'=q\right)$.
By calling $P^{ij}\left(q\right)=\eta^{i}\otimes\eta^{j}\left(\sigma\sigma'=q\right)$
we can write $P\left(q\right)$ as 
\begin{equation}
P\left(q\right)=N_{\Sigma}^{-2}\sum_{0\leq i,j\leq N_{\Sigma}}P^{ij}\left(q\right).
\end{equation}
Let $Q_{q}$ the set of vectors $\left\{ q_{a}\right\} _{1\leq a\leq n}$
, $q_{a}\in\left\{ -1,1\right\} $ such that $\sum_{a}q_{a}=q$, and
let 
\begin{equation}
P_{a}^{ij}\left(q_{a}\right)=\eta_{a}^{i}\otimes\eta_{a}^{j}\left(\sigma_{a}\sigma'_{a}=q_{a}\right)
\end{equation}
the conditional distribution marginalized over a given site $a$,
where by Eq. (\ref{eq:TAUMODEL}) holds $P_{a}^{ij}\left(1\right)=\tau_{a}^{i}\tau_{a}^{j}+\bar{\tau}_{a}^{i}\bar{\tau}_{a}^{j}$
and $P_{a}^{ij}\left(-1\right)=\tau_{a}^{i}\bar{\tau}_{a}^{j}+\bar{\tau}_{a}^{i}\tau_{a}^{j}$.
Then we can write each $P^{ij}\left(q\right)$ as 
\begin{equation}
P^{ij}\left(q\right)=\sum_{\left\{ q_{a}\right\} _{1\leq a\leq n}\in\,Q_{q}}\ \prod_{a=1}^{n}P_{a}^{ij}\left(q_{a}\right).
\end{equation}
Notice that the $P^{ij}\left(q\right)$ are multinomial distributions
parameterized by the $T$ vectors, and by simple combinatorial arguments
it is not hard to see that each of the above $q/n$ will be tight
distributed around their mean values 
\begin{equation}
q_{ij}=\frac{1}{n}\sum_{a=1}^{n}\tau_{a}^{i}\tau_{a}^{j}=N_{V}^{-1}\sum_{\alpha=1}^{N_{V}}m_{\alpha}^{i}m_{\alpha}^{j}
\end{equation}
for $n\rightarrow\infty$. It is also evident that the overlap distribution
$P\left(q\right)$ of the approximated model $\left(\Omega^{n},\eta\right)$
and that of the original one $\left(\Omega^{n},\mu\right)$ are the
same up to an error $2\epsilon$ in total variation due to the irregular
sets. 

\section{Random kernel tree}

In the previous section we presented an argument to approximate any
measure on $\Omega^{n}$ by a weighted mixture of product measures.
Here we generalize the argument by defining a ``cascade'' of nested
approximations, where each measure of the product is itself approximated
by a mixture of product measure and so on. 

Let first consider a two step approximation. In the previous section
we wrote 
\begin{equation}
\mu\left(\sigma\right)=\sum_{i}\zeta^{i_{1}}\prod_{\alpha_{1}}\eta_{\alpha_{1}}^{i_{1}}\left(\sigma\right)+\delta\left(\epsilon_{1}\right)
\end{equation}
where $\eta_{\alpha_{1}}^{i_{1}}\left(\sigma\right)$ are product
measures at the single spin level and $\delta\left(\epsilon_{1}\right)\rightarrow0$
as $\epsilon_{1}\rightarrow0$. Now instead we consider an intermediate
stage of decomposition where 
\begin{equation}
\eta_{\alpha_{1}}^{i_{1}}\left(\sigma\right)=\sum_{i_{2}}\zeta_{\alpha_{1}}^{i_{1}i_{2}}\prod_{\alpha_{2}}\eta_{\alpha_{1}\alpha_{2}}^{i_{1}i_{2}}\left(\sigma\right),
\end{equation}
the $\zeta_{\alpha_{1}}^{i_{1}i_{2}}$ are opportune weights and $\eta_{\alpha_{1}\alpha_{2}}^{i_{1}i_{2}}\left(\sigma\right)$
is a single spin product as in Eq. (\ref{eq:fundamental-1}). 
\begin{equation}
\eta_{\alpha_{1}\alpha_{2}}^{i_{1}i_{2}}\left(\sigma\right)=\prod_{a\in V_{\alpha_{1}\alpha_{2}}}\left(\frac{1+m_{\alpha_{1}\alpha_{2}}^{i_{1}i_{2}}\sigma_{a}}{2}\right)
\end{equation}
Then we can write the second step
\begin{equation}
\mu\left(\sigma\right)=\sum_{i}\zeta^{i_{1}}\prod_{\alpha_{1}}\sum_{i_{2}}\zeta_{\alpha_{1}}^{i_{1}i_{2}}\prod_{\alpha_{2}}\eta_{\alpha_{1}\alpha_{2}}^{i_{1}i_{2}}\left(\sigma\right)+\delta\left(\epsilon_{2}\right).
\end{equation}
Before generalizing the above construction to an arbitrary level of
refinement we need some new notation, ie a kernel sequence defined
trough a series of equitable refinements of $\left[0,1\right)\setminus V_{0},\left[0,1\right)\setminus S_{0}$.
Then, let $L\in\mathbb{N}$ and
\begin{equation}
X=\left\{ x_{\ell}\right\} _{0\leq\ell\leq L+1},
\end{equation}
a collection of real variables satisfying $x_{\ell+1}<x_{\ell},$
with $x_{0}=1$ and $x_{L+1}=0$, and define a sequence of equitable
refinements as follows. We start from 
\begin{equation}
(V^{0},\Sigma^{0})=(\left[0,1\right)\setminus V_{0},\left[0,1\right)\setminus S_{0})
\end{equation}
and each subsequent refinement 
\begin{equation}
(V^{\ell},\Sigma^{\ell})=(\,{\textstyle \{V_{\alpha_{1}\,...\,\alpha_{\ell}}^{\ell}\},\,\{\Sigma_{i_{1}\,...\,i_{\ell}}^{\ell}\})}
\end{equation}
is obtained by splitting each subset of $V^{\ell-1}$ into $p_{\ell}=x_{\ell-1}/x_{\ell}$
subsets of equal size $|V^{\ell}|=x_{\ell}n$ and each subset of $\Sigma^{\ell-1}$
into a number $|\Omega|^{t_{\ell}n}$, $t_{\ell}=x_{\ell-1}-x_{\ell}$
of subsets of size $|\Sigma^{\ell}|=|\Omega|^{x_{\ell}n}$. More precisely,
\begin{equation}
\begin{array}{l}
V_{\alpha_{1}\,...\,\alpha_{\ell-1}}^{\ell-1}=\bigcup_{1\leq\alpha_{\ell}\leq p_{\ell}}V_{\alpha_{1}\,...\,\alpha_{\ell-1}\alpha_{\ell}}^{\ell},\\
\\\Sigma_{i_{1}\,...i_{\ell-1}}^{\ell-1}\ =\,\bigcup_{\,1\leq i_{\ell}\leq|\Omega|^{t_{\ell}n}}\,\Sigma_{\ i_{1}\,...\,i_{\ell-1}i_{\ell}}^{\ell}.
\end{array}
\end{equation}
The last effective level $L$ of our refinements is settled to be
the equitable $\left(V,\Sigma\right)$ defined before, with $N_{V}=1/x_{L}$
and $N_{\Sigma}=|\Omega|^{(1-x_{L})n}$, while by taking $x_{L+1}=0$
we conventionally assume that the level $L+1$ is just the array $T$
of Eq. (\ref{eq:TAU}). For notation convenience we will often abbreviate
$\alpha_{1}...\alpha_{\ell}=\boldsymbol{\alpha}_{\ell}$ and $i_{1}...i_{\ell}=\boldsymbol{i}_{\ell}$.
By the properties of regularity partitions all these refinements can
be constructed to be $\epsilon_{\ell}-$regular for some strictly
decreasing real valued sequence 
\begin{equation}
\epsilon=\epsilon_{L}<...<\epsilon_{\ell}<\epsilon_{\ell-1}<...<\epsilon_{0}\leq1,
\end{equation}
and comes with a sequence of kernels 
\begin{equation}
M^{\ell}=\left\{ m_{\alpha_{1}...\alpha_{\ell}}^{i_{1}...i_{\ell}}\right\} _{1\leq\alpha_{\ell}\leq p_{\ell},\,1\leq i_{\ell}\leq\left|\Omega\right|^{t_{\ell}n}}
\end{equation}
whose entries values are given by 
\begin{equation}
m_{\boldsymbol{\alpha}_{\ell}}^{\boldsymbol{i}_{\ell}}=\overline{W}\left(V_{\alpha_{1}...\alpha_{\ell}}^{\ell},\Sigma_{i_{1}...i_{\ell}}^{\ell}\right).
\end{equation}
For the last level of refinement $L$ we will often use the notation
$M^{L}=M$ with 
\begin{equation}
M=\left\{ m^{i}\right\} _{1\leq i\leq N_{\Sigma}},\ m^{i}=\left\{ m_{\alpha}^{i}\right\} _{1\leq\alpha\leq N_{V}}.
\end{equation}
Also, we conventionally set $M^{L+1}=T$, while for the zero level
we will denote $m=\overline{W}\left(V^{0},\Sigma^{0}\right)$. Finally,
we introduce a variable which will play a central role in our construction.
Let introduce the difference between the magnetization kernels at
two consecutive levels of refinement, we indicate them with
\begin{equation}
\delta m{}_{\boldsymbol{\alpha}_{\ell}}^{\boldsymbol{i}_{\ell}}=m_{\boldsymbol{\alpha}_{\ell}}^{\boldsymbol{i}_{\ell}}-m_{\boldsymbol{\alpha}_{\ell-1}}^{\boldsymbol{i}_{\ell-1}}.
\end{equation}
By construction this variable satisfies
\begin{equation}
\sum_{\boldsymbol{\alpha}_{\ell}}\delta m{}_{\boldsymbol{\alpha}_{\ell}}^{\boldsymbol{i}_{\ell}}=0,\ |\delta m{}_{\boldsymbol{\alpha}_{\ell}}^{\boldsymbol{i}_{\ell}}|\leq\epsilon_{\ell}.
\end{equation}
We can redefine the above in terms of cavity kernels of Eq. (\ref{eq:CAVITY})
by taking 
\begin{equation}
H^{\ell}=\left\{ h_{\alpha_{1}...\alpha_{\ell}}^{i_{1}...i_{\ell}}\right\} _{1\leq\alpha_{\ell}\leq p_{\ell},\,1\leq i_{\ell}\leq\left|\Omega\right|^{t_{\ell}n}},
\end{equation}
and following the steps before. In this case we define $\delta h{}_{\boldsymbol{\alpha}_{\ell}}^{\boldsymbol{i}_{\ell}}=h_{\boldsymbol{\alpha}_{\ell}}^{\boldsymbol{i}_{\ell}}-h_{\boldsymbol{\alpha}_{\ell-1}}^{\boldsymbol{i}_{\ell-1}}$
such that 
\begin{equation}
\sum_{\boldsymbol{\alpha}_{\ell}}\delta h{}_{\boldsymbol{\alpha}_{\ell}}^{\boldsymbol{i}_{\ell}}=0,\ |\delta h{}_{\boldsymbol{\alpha}_{\ell}}^{\boldsymbol{i}_{\ell}}|<\infty.
\end{equation}
It is important to notice that the scaling of the block sizes 
\begin{equation}
|\Sigma_{i_{1}\,...\,i_{\ell}}^{\ell}|=\left|\Omega\right|^{x_{\ell}n},\ |V_{\alpha_{1}\,...\,\alpha_{\ell}}^{\ell}|=x_{\ell}n,
\end{equation}
which forces the proportions between the sizes of $V_{\alpha_{1}\,...\,\alpha_{\ell}}^{\ell}$
and $\Sigma_{i_{1}\,...\,i_{\ell}}^{\ell}$ to be
\begin{equation}
|V_{\alpha_{1}\,...\,\alpha_{\ell}}^{\ell}|\log\left|\Omega\right|=\log|\Sigma_{i_{1}\,...\,i_{\ell}}^{\ell}|
\end{equation}
for each level of refinement has been taken in order to allow the
variance of $\delta{}_{\alpha_{1}...\alpha_{\ell}}^{i_{1}...i_{\ell}}$
to be non trivial, ie of order $O\left(1\right)$, in the $n\rightarrow\infty$
limit for each level $\ell$. Again, the choice of the ratio constant
is uninfluent and has been settled equal to $\log\left|\Omega\right|$
to recover the same ratio between the sizes of $V^{L+1}=\left\{ 1,\,...,\,n\right\} $
and $\Sigma^{L+1}=\Omega^{n}$ for any pair $V^{\ell}$, $\Sigma^{\ell}$
(the implicit will would be to interpret these subsets as the ``replicas''
of the RSB ansatz).

Using the above structure we can now iterate the procedure described
at the beginning of this section to define an approximate measure
\begin{equation}
\eta\left(\sigma\right)=\sum_{i_{1}}\zeta^{\boldsymbol{i}_{1}}\prod_{\alpha_{1}}\sum_{i_{2}}\zeta_{\boldsymbol{\alpha}_{1}}^{\boldsymbol{i}_{2}}\prod_{\alpha_{2}}\,...\,\sum_{i_{L}}\zeta_{\boldsymbol{\alpha}_{L-1}}^{\boldsymbol{i}_{L}}\prod_{\alpha_{L}}\eta_{\boldsymbol{\alpha}_{L}}^{\boldsymbol{i}_{L}}\left(\sigma\right)
\end{equation}
where the last level is defined as 
\begin{equation}
\eta_{\boldsymbol{\alpha}_{L}}^{\boldsymbol{i}_{L}}\left(\sigma\right)=\sum_{i_{L}\in S_{\boldsymbol{i}_{L-1}}}\zeta_{\boldsymbol{\alpha}_{L-1}}^{\boldsymbol{i}_{L}}\prod_{a\in V_{\boldsymbol{\alpha}_{L}}}\left(\frac{1+m_{\boldsymbol{\alpha}_{L}}^{\boldsymbol{i}_{L}}\sigma_{a}}{2}\right).
\end{equation}
We can also rewrite the above using the following recursion
\begin{equation}
\eta_{\boldsymbol{\alpha}_{\ell-1}}^{\boldsymbol{i}_{\ell-1}}\left(\sigma\right)=\sum_{i_{\ell}\in S_{\boldsymbol{i}_{\ell-1}}}\zeta_{\boldsymbol{\alpha}_{\ell-1}}^{\boldsymbol{i}_{\ell}}\prod_{\alpha_{\ell}\in V_{\boldsymbol{\alpha}_{\ell-1}}}\eta_{\boldsymbol{\alpha}_{\ell}}^{\boldsymbol{i}_{\ell}}\left(\sigma\right)
\end{equation}
and if we change to the random variables as before we finally obtain
\begin{equation}
\boldsymbol{\eta}_{\boldsymbol{\alpha}_{\ell-1}}^{\boldsymbol{i}_{\ell-1}}\left(\sigma\right)=\mathbb{E}_{\boldsymbol{m}_{\boldsymbol{\alpha}_{\ell}}^{\boldsymbol{i}_{\ell}}}\prod_{\alpha_{\ell}\in V_{\boldsymbol{\alpha}_{\ell-1}}}\boldsymbol{\eta}_{\boldsymbol{\alpha}_{\ell}}^{\boldsymbol{i}_{\ell}}\left(\sigma\right)\label{eq:nesting}
\end{equation}
where the last stage is given by
\begin{equation}
\boldsymbol{\eta}_{\boldsymbol{\alpha}_{L}}^{\boldsymbol{i}_{L}}\left(\sigma\right)=\mathbb{E}_{\boldsymbol{m}_{\boldsymbol{\alpha}_{L}}^{\boldsymbol{i}_{L}}}\prod_{a\in V_{\boldsymbol{\alpha}_{L}}}\left(\frac{1+\boldsymbol{m}_{\boldsymbol{\alpha}_{L}}^{\boldsymbol{i}_{L}}\sigma_{a}}{2}\right)
\end{equation}
At this point we need some considerations. Since Eq. (\ref{eq:nesting})
is just an alternative way to rewrite Eq. (\ref{eq:aproximantmes})
by introducing a large redundant set of probability measures this
may seem just a useless complication. However as we shall see in short
this is not the case. This construction has in fact a nice physical
significance, since the levels of refinements defined above are indeed
a way to look at how fluctuations behaves while changing the sizes
of our partitions (it quite resemble a Migdal-Kadanoff renormalization
scheme on the kernel space, where the blocks are defined trough the
regularity partitions).

\section{The RSB scheme. }

The first step to obtain the RSB ansatz from the above construction
is the Gaussian approximation presented at the end of the second section.
Given that $M^{\ell}$ are doubly exchangeable arrays and that 
\begin{equation}
|\,\Sigma_{\alpha_{1}...\alpha_{\ell-1}}^{i_{1}...i_{\ell-1}}|=\left|\Omega\right|^{t_{\ell}n}\rightarrow\infty
\end{equation}
as $n\rightarrow\infty$, we could again try to invoke the Central
Limit Theorem and approximate in law $\delta m_{\alpha_{1}...\alpha_{\ell}}^{i_{1}...i_{\ell}}$
and $\delta h_{\alpha_{1}...\alpha_{\ell}}^{i_{1}...i_{\ell}}$ as
centered Gaussian random variables 
\begin{equation}
\delta\boldsymbol{m}_{\boldsymbol{\alpha}_{\ell}}^{\boldsymbol{i}_{\ell}}\overset{\mathcal{L}}{=}\boldsymbol{z}_{\boldsymbol{\alpha}_{\ell}}^{\boldsymbol{i}_{\ell}}\,\sqrt{\gamma_{\boldsymbol{\alpha}_{\ell-1}}^{\boldsymbol{i}_{\ell-1}}}\,,\label{eq:kernel=000020law}
\end{equation}
where, $\boldsymbol{z}_{\boldsymbol{\alpha}_{\ell}}^{\boldsymbol{i}_{\ell}}$
are i.i.d. standard Gaussians of unitary variance, and the scaling
parameters 
\begin{equation}
\gamma_{\boldsymbol{\alpha}_{\ell-1}}^{\boldsymbol{i}_{\ell-1}}=\mathbb{E}_{\delta\boldsymbol{m}_{\boldsymbol{\alpha}_{\ell}}^{\boldsymbol{i}_{\ell}}}|\delta\boldsymbol{m}_{\boldsymbol{\alpha}_{\ell}}^{\boldsymbol{i}_{\ell}}|^{2}\leq\epsilon_{\ell}^{2}
\end{equation}
are empirical estimators of the variance inside each block of the
$\ell-1$ level. The main idea is to discard any information on higher
cumulant and concentrate on the firs two, which clearly force us to
consider Gaussian distributions (the Gaussian distribution is the
only probability distribution with vanishing higher order cumulants).
Notice that we can think this picture as a ``cascade'' of Gaussian
measures where the random variables $\boldsymbol{m}_{\boldsymbol{\alpha}_{\ell-1}}^{\boldsymbol{i}_{\ell-1}}$
are by themselves controlling the distribution of the $\ell-$th level
\begin{equation}
\boldsymbol{m}_{\boldsymbol{\alpha}_{\ell}}^{\boldsymbol{i}_{\ell}}\thicksim N\left(\boldsymbol{m}_{\boldsymbol{\alpha}_{\ell-1}}^{\boldsymbol{i}_{\ell-1}},\gamma_{\boldsymbol{\alpha}_{\ell-1}}^{\boldsymbol{i}_{\ell-1}}\right),
\end{equation}
It's easy to make the above recursion explicit and rewrite $\boldsymbol{m}_{\boldsymbol{\alpha}_{L}}^{\boldsymbol{i}_{L}}$
as a sum
\begin{equation}
\boldsymbol{m}_{\boldsymbol{\alpha}_{L}}^{\boldsymbol{i}_{L}}=m+\sum_{\ell=1}^{L}\boldsymbol{z}_{\boldsymbol{\alpha}_{\ell}}^{\boldsymbol{i}_{\ell}}\sqrt{\gamma_{\boldsymbol{\alpha}_{\ell-1}}^{\boldsymbol{i}_{\ell-1}}},\ \boldsymbol{z}_{\boldsymbol{\alpha}_{\ell}}^{\boldsymbol{i}_{\ell}}\thicksim N\left(0,1\right)
\end{equation}
where the $\boldsymbol{z}_{\boldsymbol{\alpha}_{\ell}}^{\boldsymbol{i}_{\ell}}$
are i.i.d standard Gaussians. Also, since $\partial_{s}\tanh^{-1}\left(s\right)|_{s=0}=1$,
then 
\begin{equation}
\boldsymbol{h}_{\boldsymbol{\alpha}_{\ell}}^{\boldsymbol{i}_{\ell}}-h\overset{\mathcal{L}}{=}\,\boldsymbol{m}_{\boldsymbol{\alpha}_{\ell-1}}^{\boldsymbol{i}_{\ell-1}}-m,
\end{equation}
and we can represent also the $\boldsymbol{h}_{\boldsymbol{\alpha}_{\ell}}^{\boldsymbol{i}_{\ell}}$
in distribution as a random variables with 
\begin{equation}
\boldsymbol{h}_{\boldsymbol{\alpha}_{\ell}}^{\boldsymbol{i}_{\ell}}\,\overset{\mathcal{L}}{=}\,h+\sum_{1\leq\ell\leq L}\boldsymbol{z}_{\boldsymbol{\alpha}_{\ell}}^{\boldsymbol{i}_{\ell}}\,\sqrt{\gamma_{\boldsymbol{\alpha}_{\ell-1}}^{\boldsymbol{i}_{\ell-1}}}.
\end{equation}
To resume, and given that $n$ is large, our aim is to approximate
in law the randomly mixed kernel by a sequence of refinements $(V^{\ell},\Sigma^{\ell})$,
an offset and the arrays 
\begin{equation}
\Gamma=\left\{ \Gamma^{\ell}\right\} _{1\leq\ell\leq L},\ \Gamma^{\ell}=\left\{ \gamma_{\alpha_{1}...\alpha_{\ell}}^{i_{1}...i_{\ell}}\right\} _{1\leq\alpha_{\ell}\leq p_{\ell},\,1\leq i_{\ell}\leq\left|\Omega\right|^{t_{\ell}n}}\label{eq:GAMMONE}
\end{equation}
using of Eq. (\ref{eq:kernel=000020law}). By construction the above
array sequence is hierarchically organized, since each block of variables
of the $\ell-$th level is controlled by a law which only depends
on a block of the $\ell-1$ level. Assuming this Gaussian approximation
we can rewrite $\eta\left(\sigma\right)$ by the recursive formula
\begin{equation}
\boldsymbol{\eta}_{\boldsymbol{\alpha}_{\ell-1}}^{\boldsymbol{i}_{\ell-1}}\left(\sigma\right)=\mathbb{E}_{\boldsymbol{z}_{\boldsymbol{\alpha}_{L}}^{\boldsymbol{i}_{L}}}\prod_{\alpha_{\ell}\in V_{\boldsymbol{\alpha}_{\ell-1}}}\boldsymbol{\eta}_{\boldsymbol{\alpha}_{\ell}}^{\boldsymbol{i}_{\ell}}\left(\sigma\right)\label{eq:recursion66-1}
\end{equation}
starting from the bottom level
\begin{equation}
\boldsymbol{\eta}_{\boldsymbol{\alpha}_{L}}^{\boldsymbol{i}_{L}}\left(\sigma\right)=\prod_{a\in V_{\boldsymbol{\alpha}_{\ell-1}}}\left[\frac{1+(m+\sum_{\ell}\boldsymbol{z}_{\boldsymbol{\alpha}_{\ell}}^{\boldsymbol{i}_{\ell}}\,{\scriptstyle \sqrt{\gamma_{\boldsymbol{\alpha}_{\ell-1}}^{\boldsymbol{i}_{\ell-1}}}}\,)\,\sigma_{a}}{2}\right].
\end{equation}
It seems that even with this simplification we didn't solve much,
since we still have a bounch of free parameters which grows exponentially
in $L$. But notice that if we assume that the fluctuations have the
same amplitude for each level, ie $\gamma_{\boldsymbol{\alpha}_{\ell-1}}^{\boldsymbol{i}_{\ell-1}}\rightarrow\gamma_{\ell}$,
then the recursive formula in Eq. (\ref{eq:recursion66-1}) becomes
\begin{equation}
\boldsymbol{\eta}_{\boldsymbol{\alpha}_{\ell-1}}^{\boldsymbol{i}_{\ell-1}}\left(\sigma\right)=\mathbb{E}_{\boldsymbol{z}_{\boldsymbol{\alpha}_{L}}^{\boldsymbol{i}_{L}}}\left[\,\boldsymbol{\eta}_{\boldsymbol{\alpha}_{\ell}}^{\boldsymbol{i}_{\ell}}\left(\sigma\right)^{p_{\ell}}\right]\label{eq:recursionRSB}
\end{equation}
with $p_{\ell}=x_{\ell-1}/x_{\ell}$ and the last level given by 
\begin{equation}
\boldsymbol{\eta}_{\boldsymbol{\alpha}_{L}}^{\boldsymbol{i}_{L}}\left(\sigma\right)=\prod_{a\in V_{\boldsymbol{\alpha}_{\ell-1}}}\left[\frac{1+\left(m+\sum_{\ell}\boldsymbol{z}_{\boldsymbol{\alpha}_{\ell}}^{\boldsymbol{i}_{\ell}}\,\sqrt{\gamma_{\ell}}\right)\sigma_{a}}{2}\right],
\end{equation}
which, as we shall see, is exactly the $L-$th level RSB measure we
are searching for. This constraint dramatically reduces the number
of parameters, and is expected to be equivalent to the ``overlap
equivalence'' assumption \cite{Parisi=000020ReEquiv}. To deal with
a more familiar expression we can rewrite the above measure using
the cavity kernels 
\begin{equation}
\boldsymbol{\eta}_{\boldsymbol{\alpha}_{L}}^{\boldsymbol{i}_{L}}\left(\sigma\right)=\prod_{a\in V_{\boldsymbol{\alpha}_{\ell-1}}}\frac{1}{\boldsymbol{Z}_{\boldsymbol{\alpha}_{L}}^{\boldsymbol{i}_{L}}}e^{-\left(h+\sum_{\ell}\boldsymbol{z}_{\boldsymbol{\alpha}_{\ell}}^{\boldsymbol{i}_{\ell}}\,\sqrt{\gamma_{\ell}}\right)\sigma_{a}},\label{eq:cavitymeasure}
\end{equation}
where the normalizations are given by
\begin{equation}
\log\boldsymbol{Z}_{\boldsymbol{\alpha}_{L}}^{\boldsymbol{i}_{L}}=\log2+\log\cosh\left({\textstyle h+\sum_{\ell}\boldsymbol{z}_{\boldsymbol{\alpha}_{\ell}}^{\boldsymbol{i}_{\ell}}\,\sqrt{\gamma_{\ell}}}\right).
\end{equation}
If we rewrite Eq. (\ref{eq:recursionRSB}) by defining the auxiliary
variable $\boldsymbol{U}_{\ell}^{x_{\ell}}=\boldsymbol{\eta}_{\boldsymbol{\alpha}_{\ell}}^{\boldsymbol{i}_{\ell}}$
we obtain
\begin{equation}
\boldsymbol{U}_{\ell-1}^{x_{\ell-1}}=\mathbb{E}_{\boldsymbol{z}_{\boldsymbol{\alpha}_{L}}^{\boldsymbol{i}_{L}}}\left(\boldsymbol{U}_{\ell}^{x_{\ell-1}}\right)\label{eq:recursion66}
\end{equation}
and we can recognize the recursion to obtain the cavity part of the
Parisi formula (if we take $\gamma_{\ell}=q_{\ell+1}-q_{\ell}$ as
in the usual notation, see next section). It is quite long but not
hard at all to prove that if $\gamma_{\boldsymbol{\alpha}_{\ell-1}}^{\boldsymbol{i}_{\ell-1}}=\gamma_{\ell}$
is assumed these weights come indeed from a Ruelle Cascade \cite{Bolthausen},
however we demand this to a future work and spend the rest of this
paper in obtaining the Parisi formula for the SK model. 

\section{The SK model.}

We are now ready to apply our considerations to the the Sherrington-Kirkpatrick
(SK) model. For our convenience we only consider the case without
the external magnetic field. The SK model is described by the random
Hamiltonian 
\begin{equation}
\mathcal{H}_{\boldsymbol{J}}\left(\sigma\right)=\frac{1}{2\sqrt{n}}\sum_{1\leq a,b\leq n}\boldsymbol{J}_{ab}\sigma_{a}\sigma_{b},
\end{equation}
where $\boldsymbol{J}=\left\{ \boldsymbol{J}_{ab}\right\} _{1\leq a,b\leq n}$
is a symmetric random matrix with Gaussian entries such that $\boldsymbol{J}_{ab}=\boldsymbol{J}_{ba}$,
$\boldsymbol{J}_{aa}=0$, and such that the average is zero and the
variance one,
\begin{equation}
\mathbb{E}_{\boldsymbol{J}}(\boldsymbol{J}_{ab})=0,\ \mathbb{E}_{\boldsymbol{J}}(\boldsymbol{J}_{ab}^{2})=1.
\end{equation}
The main task is to compute the averaged free energy
\begin{equation}
\bar{f}=\lim_{n\rightarrow\infty}n^{-1}\mathbb{E}_{\boldsymbol{J}}\log Z_{\boldsymbol{J}},
\end{equation}
which is provided by the celebrated formula by Parisi. Let $L$ be
the number of RSBs and take two real sequences $\{x_{\ell}\}$ and
$\left\{ q_{\ell}\right\} $ such that $x_{\ell-1}\leq x_{\ell}$,
$q_{\ell-1}\leq q_{\ell}$, $x_{0}=q_{0}=0$ and $x_{L+1}=q_{L+1}=1$.
Now let
\begin{equation}
\boldsymbol{Y}_{L+1}=\cosh\left(\beta\,{\textstyle \sum_{\ell=0}^{L}\,}\boldsymbol{z}_{\ell}\,\sqrt{q_{\ell+1}-q_{\ell}}\right),
\end{equation}
where $\boldsymbol{z}_{\ell}$ are i.i.d. standard Gaussian random
variables of unitary variance, and iterate
\begin{equation}
\boldsymbol{Y}_{\ell-1}^{x_{\ell-1}}=\mathbb{E}_{\boldsymbol{z}_{\ell}}\left(\boldsymbol{Y}_{\ell}^{x_{\ell-1}}\right)\label{eq:iteratePArisi}
\end{equation}
up to $Y_{1}$. Then the Parisi functional is 
\begin{equation}
f_{L}=\log2+\log Y_{1}-\frac{\beta^{2}}{2}\sum_{\ell\geq1}x_{\ell}\left(q_{\ell+1}^{2}-q_{\ell}^{2}\right)\label{eq:Parisiformula}
\end{equation}
Notice that in case of zero external field no randomness remains after
the iterations of Eq. (\ref{eq:iteratePArisi}), and the free energy
is self-averaging with respect to the disorder. It has been argued
by Parisi, then proved by Guerra and Talagrand, that 
\begin{equation}
\bar{f}=\inf_{L,\{x_{\ell}\},\{q_{\ell}\}}f_{L}.
\end{equation}
We will show that the measure $\eta\left(\sigma\right)$ in Eq.s (\ref{eq:recursion66})
and (\ref{eq:cavitymeasure}) can be applied to the Aizenmann-Simm-Starr
(ASS) cavity representation of the free energy to obtain the above
formula. The ASS representation of the SK free energy is \cite{Bolthausen}
\begin{equation}
A_{n}=\log2+\mathbb{E}_{\boldsymbol{J}}\log\langle\cosh\left(\beta\boldsymbol{y}_{n+1}\left(\sigma\right)\right)\rangle_{\boldsymbol{G}}-\mathbb{E}_{\boldsymbol{J}}\log\langle\exp(\beta\boldsymbol{\kappa}\left(\sigma\right)/\sqrt{2})\rangle_{\boldsymbol{G}}\label{eq:ASS}
\end{equation}
where $\langle\,\cdot\,\rangle_{\boldsymbol{G}}$ denotes the average
with respect to the Gibbs measure $\boldsymbol{G}$ of Hamiltonian
\begin{equation}
\mathcal{H}{}_{\boldsymbol{J}}=\frac{\beta^{*}}{2\sqrt{n}}\sum_{1\leq a<b\leq n}\boldsymbol{J}_{ab}\sigma_{a}\sigma_{b},\:\beta^{*}=\beta{\textstyle \sqrt{\frac{n}{n+1}}},
\end{equation}
the $\boldsymbol{y}_{n+1}\left(\sigma\right)$ is the cavity variable
for the $n+1$ system, 
\begin{equation}
\boldsymbol{y}_{n+1}\left(\sigma\right)=\frac{1}{\sqrt{n}}\sum_{a=1}^{n}\boldsymbol{J}_{a,n+1}\sigma_{a},
\end{equation}
where the $\{\boldsymbol{J}_{a,n+1}\}$ is an additional set of i.i.d.
standard Gaussians, and $\boldsymbol{\kappa}\left(\sigma\right)$
are i.i.d. Gaussian random variables with covariance given by \cite{Bolthausen}
\begin{equation}
\mathbb{E}\left[\boldsymbol{\kappa}\left(\sigma\right)\boldsymbol{\kappa}\left(\sigma'\right)\right]=\frac{\mathbb{E}\left[\,\boldsymbol{y}_{n+1}\left(\sigma\right)\boldsymbol{y}_{n+1}\left(\sigma'\right)\right]^{2}}{2}.\label{eq:correction}
\end{equation}
We start from rewriting the Hamiltonian as 
\begin{equation}
\frac{\beta^{*}}{2\sqrt{n}}\sum_{1\leq a<b\leq n}\boldsymbol{J}_{ab}\sigma_{a}\sigma_{b}=\beta^{*}\sum_{a=1}^{n}\boldsymbol{\nu}_{a}\left(\sigma\right)\sigma_{a}
\end{equation}
where we introduced the cavity fields
\begin{equation}
\boldsymbol{\nu}_{a}\left(\sigma\right)=\frac{1}{2\sqrt{n}}\sum_{b=1}^{n}\boldsymbol{J}_{ab}\sigma_{a},\ \boldsymbol{y}_{n+1}\left(\sigma\right)\overset{\mathcal{L}}{=}\sum_{a=1}^{n}\boldsymbol{\nu}_{a}\left(\sigma\right).
\end{equation}
That given, we first concentrate on the cavity part of the ASS formula.
By the above manipulations we can use the expression 
\begin{equation}
\langle\cosh\left(\beta\boldsymbol{y}_{n+1}\left(\sigma\right)\right)\rangle_{\boldsymbol{G}}=\langle\cosh\left({\textstyle \beta\sum_{a}\boldsymbol{\nu}_{a}\left(\sigma\right)}\right)\rangle_{\boldsymbol{G}}
\end{equation}
and introduce the cavity kernel variables by rewriting $\boldsymbol{\nu}_{a}\left(\sigma\right)$
as follows 
\begin{multline}
\boldsymbol{\nu}_{a}\left(\sigma\right)=\nu_{a}^{i_{1}}+(\nu_{a}^{i_{1}i_{2}}-\nu_{a}^{i_{1}})+(\nu_{a}^{i_{1}i_{2}i_{3}}-\nu_{a}^{i_{1}i_{2}})+...\\
...+(\nu_{a}^{i_{1}...i_{\ell}}-\nu_{a}^{i_{1}...i_{\ell-1}})+...+(\boldsymbol{\nu}_{a}\left(\sigma\right)-\nu_{a}^{i_{1}...i_{L+1}}).\label{eq:dfzdfd}
\end{multline}
If we neglect the fluctuations of $\boldsymbol{\nu}_{a}\left(\sigma\right)-\nu_{a}^{i_{1}...i_{L+1}}$
by means of the kernel argument then we arrive to the expression 
\begin{equation}
\langle\cosh\left(\beta\boldsymbol{y}_{n+1}\left(\sigma\right)\right)\rangle_{\boldsymbol{G}}=\langle\cosh\left({\textstyle \beta\sum_{a}\boldsymbol{\nu}_{a}\left(\sigma\right)}\right)\rangle_{\eta}+O\left(\epsilon\right),
\end{equation}
and by assuming the RSB scheme as presented in the previous section
we can compute the above quantity by defining
\begin{equation}
\boldsymbol{Y}_{L}=\cosh\left(\beta\,{\textstyle \sum_{\ell=0}^{L}\,}\boldsymbol{z}_{\boldsymbol{\alpha}_{\ell}}^{\boldsymbol{i}_{\ell}}\,\sqrt{\gamma_{\ell-1}}\right),
\end{equation}
then iterate the formula of Eq. (\ref{eq:recursion66}) 
\begin{equation}
\boldsymbol{Y}_{\ell-1}^{x_{\ell-1}}=\mathbb{E}_{\boldsymbol{z}_{\boldsymbol{\alpha}_{\ell}}^{\boldsymbol{i}_{\ell}}}\left(\boldsymbol{Y}_{\ell}^{x_{\ell-1}}\right)
\end{equation}
and finally put this expression into the ASS representation to obtain
\begin{equation}
\langle\cosh\left({\textstyle \beta\sum_{a}\boldsymbol{\nu}_{a}\left(\sigma\right)}\right)\rangle_{\eta}=\log Y
\end{equation}
as in Eq. (\ref{eq:Parisiformula}) if we redefine $\gamma_{\ell}=q_{\ell+1}-q_{\ell}$.
The correction is computed in the very same way by rewriting the $\boldsymbol{\kappa}\left(\sigma\right)$
variables in therms of the $\boldsymbol{h}_{\boldsymbol{\alpha}_{\ell}}^{\boldsymbol{i}_{\ell}}$
according to Eq. (\ref{eq:correction}) and then computing the average
respect to the measure $\eta$ as before.

\section{Acknowledgments}

I wish to thank Amin Coja-Oghlan, Nicola Kistler, Giorgio Parisi,
Pietro Caputo and Riccardo Balzan for interesting discussions and
suggestions. This research was funded by the European Research Council
under the European Union\textquoteright s Seventh Framework Programme
(FP/2007-2013) / ERC Grant Agreement n. 278857\textendash PTCC. 

~\\
{[}\textit{NDR: This paper has been written in the period 2015-2016
within the PTCC project (Coja-Oghlan) funded by the European Research
Council}{]}
\end{document}